\documentclass[runningheads,a4paper]{llncs}

\usepackage{amssymb,amsmath}
\usepackage{graphicx}
\usepackage{url}
\usepackage{booktabs}
\urldef{\mailsa}\path{naldi@disp.uniroma2.it}
\newcommand{\keywords}[1]{\par\addvspace\baselineskip
\noindent\keywordname\enspace\ignorespaces#1}

\begin{document}

\mainmatter  

\title{Concentration in the mobile operating systems market}

\titlerunning{Mobile OS concentration}

%
%
\author{Maurizio Naldi}
\authorrunning{M. Naldi}

\institute{Universit\`{y} of Rome Tor Vergata\\Department of Computer Science and Civil Engineering\\
Via del Politecnico 1, 00133 Roma, Italy\\
\mailsa\\
}

%
%

\maketitle

\begin{abstract}
Concentration phenomena concern the ICT market. Though the regulatory action has been active mainly in the telecom network operators industry, even more significant worldwide concentration phenomena affect other industries. The market of mobile operating systems is analysed through two concentration indices to get a quantitative picture of the current situation and its evolution over time: the Hirschman Herfindahl Index (HHI) and the Four-Firm Concentration Ratio (CR4). A strongly imbalanced oligopoly is shown to exist, where the four major operating systems take over 99\% of the market, but the dominant operating system Android alone is installed on over 80\% of the new devices.  
\end{abstract}

\section{Introduction}
Market structure and the presence of dominant operators (manufacturers and/or service providers) has been a significant field of activity in industrial policy since long \cite{zlinkoff1944monopoly}. An operator holding a very large share of the market, or even acting as a monopolist, may take advantage of its position and enforce unfair policies towards its customers, which in turn have little or no room to oppose. The attention for the appearance of dominant positions is at the root of the birth of a number of national anti-trust agencies, both at the national and supernational level \cite{Alexis1995}, which enforce rules against anticompetitive agreements, abuses of dominant position as well as concentrations (e.g., mergers and acquisitions, joint ventures) which may create or strengthen dominant positions detrimental to competition.

The issue is particularly delicate in ICT industries, where operators may often benefit of economies of scale, which would lead to a natural monopolistic structure as the most efficient one \cite{nooteboom1992information}. Noam has carried out a broad analysis of concentration phenomena in several ICT and ICT-related industries \cite{Noamtprc} \cite{noam2015owns}:
\begin{itemize}
\item Books
\item Film
\item ISP
\item Magazines
\item Multi- channel
\item Newspapers
\item Online News
\item Radio
\item Search Engines
\item TV
\item Wireless
\item Wireline
\end{itemize}

In that survey, the highest HHI value is observed for search engines and is roughly 0.75, quite above the second highest value, which is 0.55 and pertains to the wireline telco market.

However, the survey of \cite{Noamtprc} leaves out a market that has often been at the center of anti-trust disputes in recent years, which is the operating systems one. The most notable ones have been the U.S.A. vs Microsoft case for the Windows desktop operating system \cite{Microsoftantitrust}, and the very recent Statement of Objections raised by the EU vs Google for the mobile operating system Android \cite{Googleantitrust}.

In that Statement of Objections, the European Commission alleges that Google has breached EU antitrust rules by:
\begin{itemize}
\item requiring manufacturers to pre-install Google Search and Google's Chrome browser and requiring them to set Google Search as default search service on their devices, as a condition to license certain Google proprietary apps;
\item preventing manufacturers from selling smart mobile devices running on competing operating systems based on the Android open source code;
\item giving financial incentives to manufacturers and mobile network operators on condition that they exclusively pre-install Google Search on their devices.
\end{itemize}

However, in both cases just brief quantitative data are given about the market structure. 

Very few scientific papers have been devoted instead to competition in the operating systems market. The effects of Microsoft Windows dominance on possible competitors and the run to standardization has been analysed in \cite{kretschmer2004upgrading}. In \cite{casadesus2006dynamic} the case of a dominant player (Windows) competing with a zero-price player (Linux) has been considered. An analysis of the desktop/laptop market for operating systems using an organizational ecology approach has instead been carried out in \cite{lakka2013competitive}, where a Lotka-Volterra model has been employed to investigate the evolution of competition and forecast equilibrium states, in the light of the presence of open source software.

The relevant literature exhibits a gap of quantitative studies concerning mobile operating systems. 

This paper provides a few data to analyse the mobile operating systems market and its evolution through the 2007-2015 period. Two concentration indicators (the Herfindahl-Hirschman Index - HHI - and the Four-firm Concentration Ratio - CR4) are employed to provide a quantitative indication of market concentration. It is shown that the dominance by Android has replaced that  by Symbian (with the cross occurring in 2011) and that the current market structure is closer to a monopoly than to a tight oligopoly (with the dominant OS holding more than 80\% of the market and the four largest OSs' share in excess of 99\%).

The paper is organized as follows. The rough data concerning the OSs' shares are reported in Section \ref{sec:shares}, while the concentration indices are computed in Section \ref{sec:hhi}.

\section{Mobile operating systems market 2007-2015}
\label{sec:shares}
In this section we provide information on the data employed to carry out the concentration analysis.

We use the raw data as provided by the quarterly market analyses released by Gartner (Gartner smartphone market share). In particular, we consider the worldwide smartphone sales, arranged by operating system, in the 2007-2015 period. The numbers refer therefore to new devices and not to the overall number of actually working phones.

The operating systems expressly considered are reported hereafter; those not appearing are cumulated as \textit{others}:
\begin{itemize}
 \item Windows Mobile;
 \item RIM;
 \item Symbian;
 \item iOS;
 \item Android,
 \item Bada;
 \item  Windows Phone. 
 \end{itemize} 

\begin{figure}
\begin{center}
\includegraphics[scale=0.5]{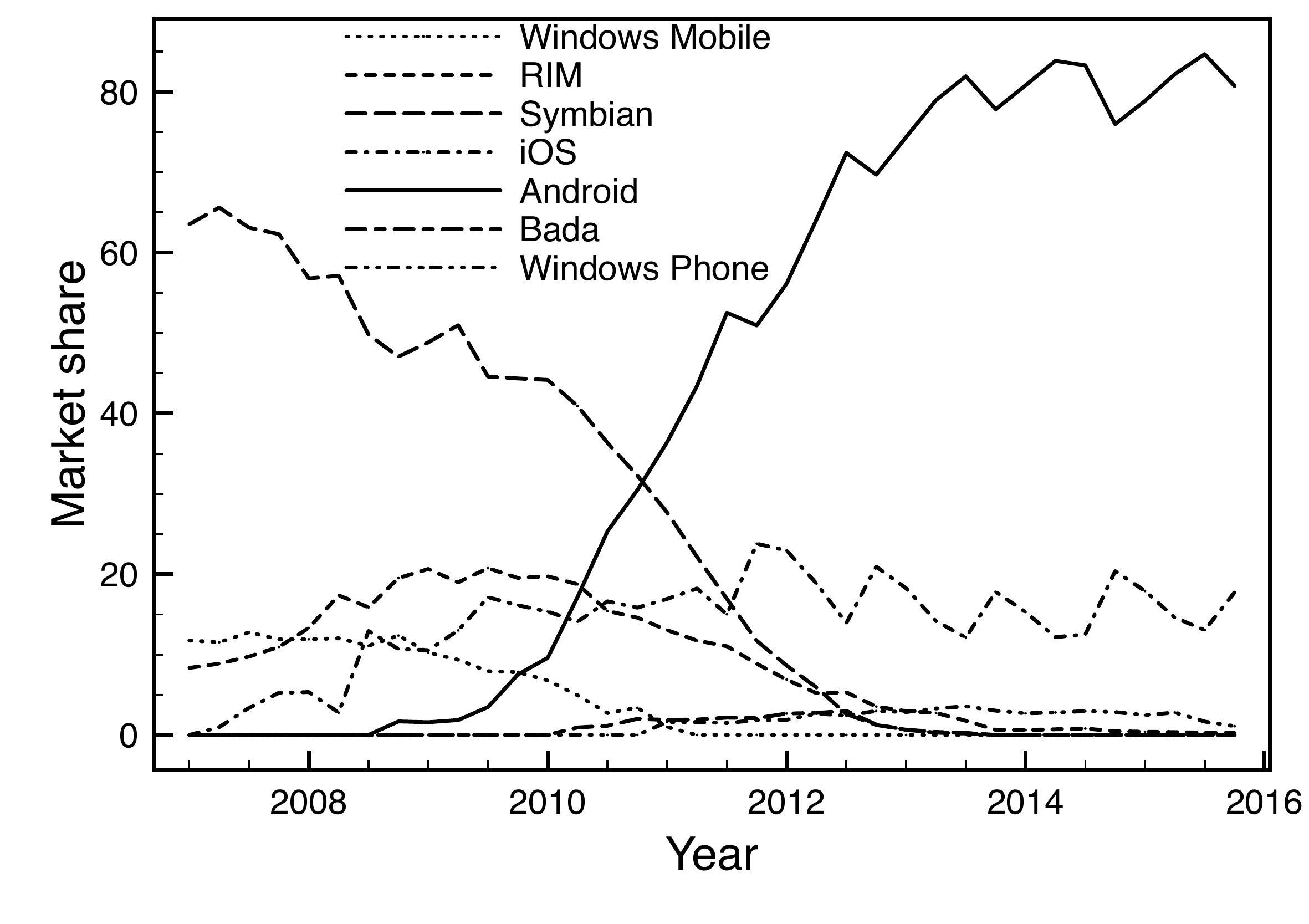}
\caption{Market share of operating systems 2007-2015}
\label{fig:tutte}
\end{center}
\end{figure}

The respective market shares (as computed over the total number of smartphone sold in the same period) are shown in \figurename~\ref{fig:tutte}

As can be seen, the evolution over the 9 year-interval can be roughly divided into two periods: the first one, where the dominance of Symbian has been steadily declining (till the negligible market share held today), and the second one where the dominance of Android has steadily risen, till a plateau reached in 2014, which is about the 80\% share of today. The crossing point, where Android has taken the lead, can be roughly identified in 2011. It is to be noted that the share held by Android today is much larger than that held by Symbian in the days of its dominance.

\section{Market concentration index}
\label{sec:hhi}
In order to assess the market structure, in this section we compute two concentration indices for the raw data mentioned in Section \ref{sec:shares}. In addition, we analyze the relevance of the operating system holding the maximum share.

As concentration indices, we consider the two most relevant indices employed in industry studies: the Herfindahl-Hirschman Index (HHI) and the Four-firm Concentration Ratio (CR4).

The HHI of a market where $n$ companies operate, and the market share of the $i$-th largest company is $s_{i}$, is 
\begin{equation}
\textrm{HHI} = \sum_{i=1}^{n}s_{i}^{2}.
\end{equation}
Its value ranges in the $[\frac{1}{n},1]$ interval, with the minimum value representing the perfect competition case (all companies having the same share of the market) and the maximum value (1) representing the absolute monopoly case: larger values represent increasing level of market concentration. Its relationship with the Zipf law, a well-known rank-size model, has been described in \cite{Naldi2003}.

In our case, we consider operating systems instead of companies.

Unfortunately, we do not have a complete picture of the market composition, since we know market shares just for the more widespread OSs. In such a case, we can however compute bounds for the HHI. Here we employ the formulas proposed in \cite{Naldi14}  and reported in Table \ref{table:fasciahhi}, where $M$ is the smallest OS whose market share we know, $R =1-\sum_{i=1}^{M}s_i$, and $Q=\lfloor R/s_M \rfloor $. 

\begin{table}
\begin{center}
\begin{tabular}{ll}
\toprule
Type & Bound\\
\midrule
Lower & $ \sum_{i=1}^{M}s_{i}^{2}$\\
Upper ($R \le s_M$) & $\sum_{i=1}^{M}s_{i}^{2} + \left( 1-\sum_{i=1}^{M}s_{i} \right)^{2}$\\
Upper ($R > s_M$) & $\sum_{i=1}^{M}s_{i}^{2} + s_M^2 Q + \left(
1-\sum_{i=1}^{M}s_{i}- s_M Q\right)^{2}$\\
\bottomrule
\end{tabular}
\end{center}\caption{Bounds on HHI}
\label{table:fasciahhi}
\end{table}

\begin{figure}
\begin{center}
\includegraphics[scale=0.5]{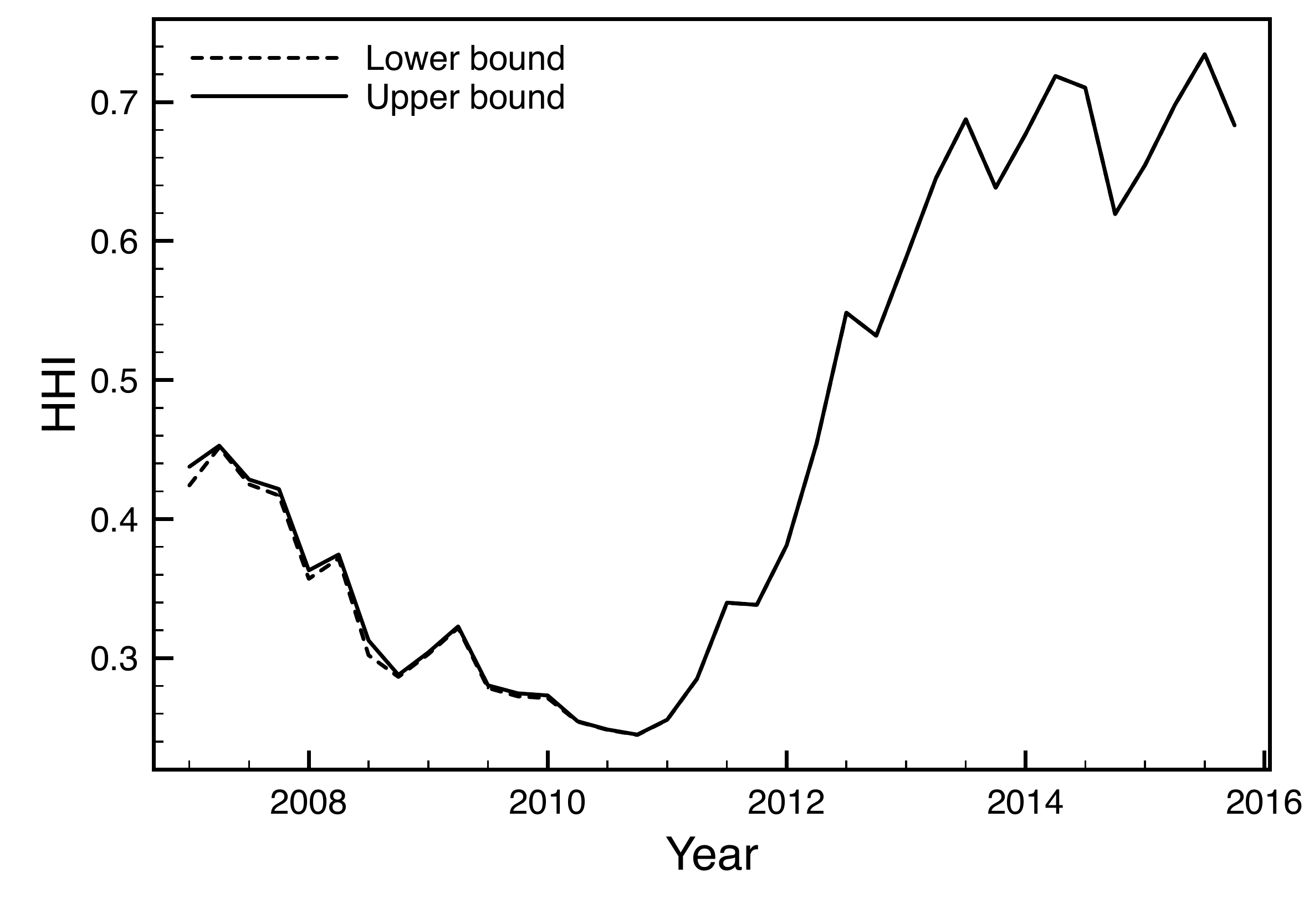}
\caption{HHI in the mobile operating systems market 2007-2015}
\label{fig:hhi}
\end{center}
\end{figure}

When we apply the formulas of Table \ref{table:fasciahhi}  to the data of Section \ref{sec:shares}, we get the curves shown in \figurename~\ref{fig:hhi}. We note that the bounds are very tight, since the two curves are practically indistinguishable: the difference between upper and lower bound is at most 3.5\% and below 1\% in 89\% of the quarters examined.

The HHI curve shows anyway that after a slight dip in 2011 (corresponding to the crossing point, where Symbian and Android held a roughly equal share of the market), the concentration indicator has steadily risen, reaching peaks in excess of 0.7. Is this a high value? Though a precise correspondence cannot be drawn between the numerical value of the HHI and the qualitative indication of a level of concentration (or, equivalently, of competition), we can resort to the guidelines provided by the U.S. Department of Justice for horizontal mergers (first in 1985 and later revised several times). In Section 5.3   of the latest version of 2010, a classification of markets into three types is proposed, as reported in Table \ref{table:hhilevel} \cite{Justice}. According to this classification, the HHI levels observed at any time during the 9-year long interval considered, and in particular in the latest years, are typical of a highly concentrated market. Actually the current value, exceeding 0.7, is by far above the threshold (0.25) that separates moderately concentrated markets from highly concentrated ones.
\begin{table}
\begin{center}
\begin{tabular}{cl}
\toprule
HHI & Competition level\\
\midrule
$<$0.15	& Unconcentrated Markets\\
0.15--0.25	& Moderately Concentrated Markets\\
$>$0.25	& Highly Concentrated Markets\\
\bottomrule\\
\end{tabular}
\caption{Levels of competition and the HHI}
\label{table:hhilevel}
\end{center}
\end{table}  

In addition to the HHI, we consider another well known concentration index: the Four-Firm Concentration Ratio (CR4). The $\mathrm{CR_4}$ index has been the most relevant index to measure concentration before the advent of the HHI \cite{Weinstock1982}. Its relationship to the HHI has been investigated in \cite{naldi2014correlation,naldi2014cr4}. It is given by the sum of the market shares of the four largest firms in the market \cite{arnold2008}
\begin{equation}
\mathrm{CR_{4}} = \sum_{i=1}^{4}s_{i}.
\end{equation}

While it is clear that a low value of the index represents a larger competition level, and a high value (close to 100) represents an oligopoly situation, there is  not a general consensus on the correspondence between the value of the index and intermediate concentrations. Typically, if $\mathrm{CR_{4}} < 0.4$, the industry is considered as very competitive. A complete classification table is proposed in \cite{Gwin} and reported here in Table \ref{table:cr4level}.
\begin{table}
\begin{center}
\begin{tabular}{cl}
\toprule
$\mathrm{CR}_{4}$ & Competition level\\
\midrule
0	& Perfect Competition\\
0--0.4	& Effective Competition or Monopolistic Competition\\
0.4--0.6 & Loose Oligopoly or Monopolistic Competition\\
$>$0.6	& Tight Oligopoly or Dominant Firm with a Competitive Fringe\\
\bottomrule\\
\end{tabular}
\caption{Levels of competition and the $\mathrm{CR_{4}}$}
\label{table:cr4level}
\end{center}
\end{table} 

\begin{figure}
\begin{center}
\includegraphics[scale=0.5]{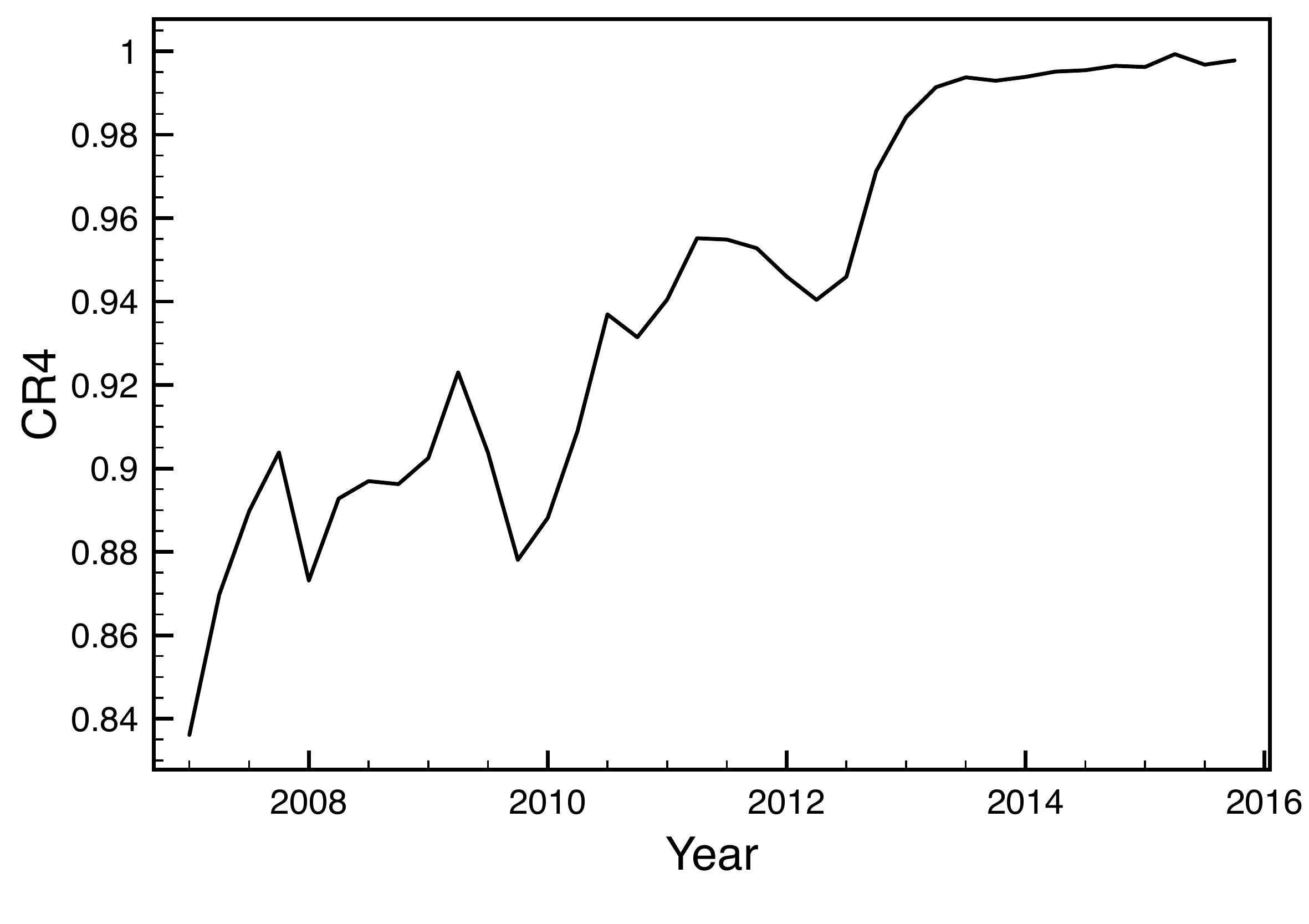}
\caption{CR4 index for the mobile operating systems market 2007-2015}
\label{fig:cr4mos}
\end{center}
\end{figure}

We show in \figurename~\ref{fig:cr4mos} the CR4 for the same data for which we computed the HHI. We observe that even in 2007 the CR4 was representative of a tight oligopoly situation, but the following years have seen a steady rise in concentration (with the slight dip again around 2011), and since 2013 the value is extremely close to 1, which means that the 4 dominant firms take all the market. Even in the years of HHI depression, the CR4  was well above the 0.6 barrier, with most of the market just redistributing among the dominant players. 

\begin{figure}
\begin{center}
\includegraphics[scale=0.5]{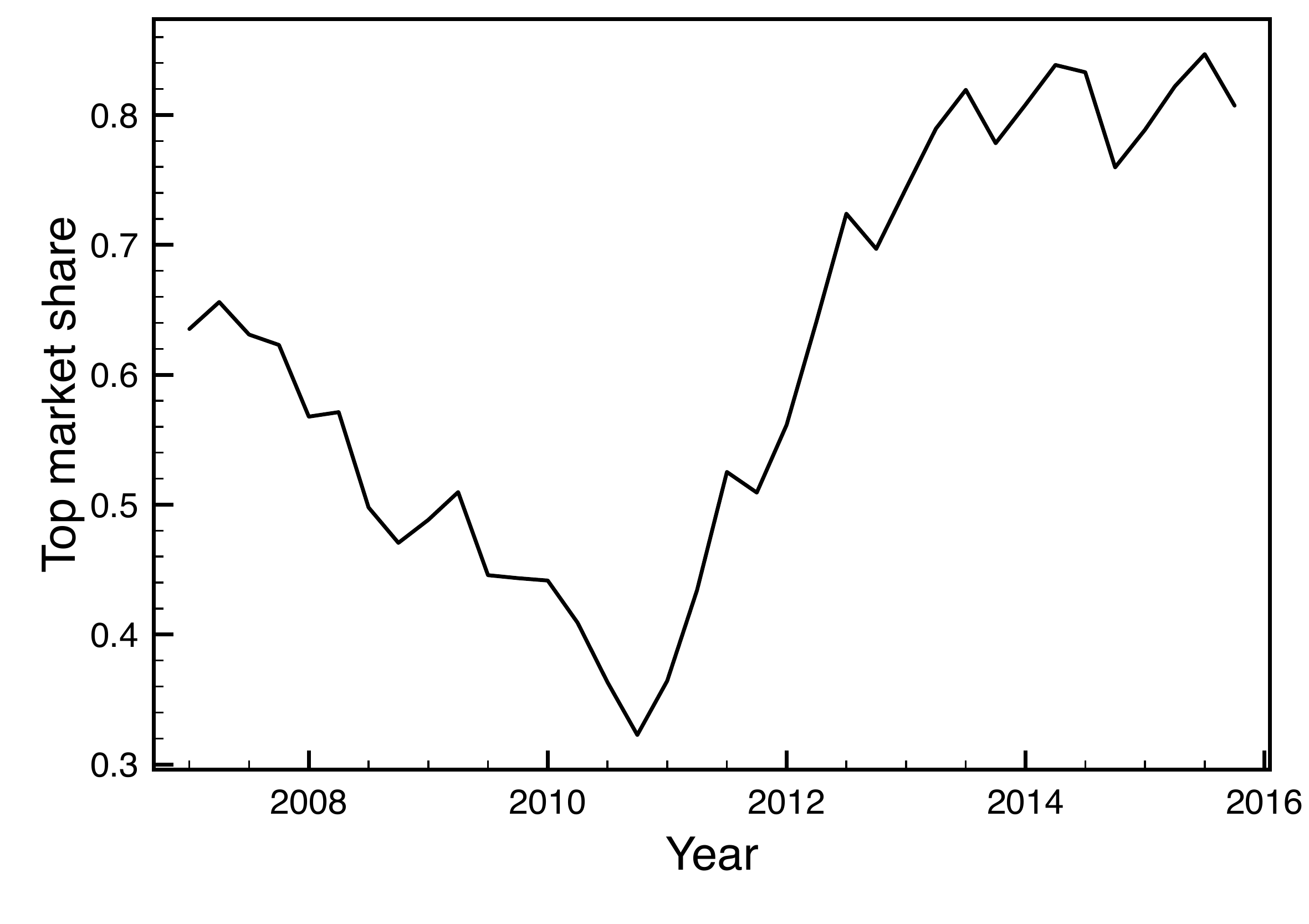}
\caption{Market share of the dominant operating system 2007-2015}
\label{fig:topos}
\end{center}
\end{figure}

Finally, in order to assess whether the market structure is closer to an oligopoly or to a monopoly, we plot in \figurename~\ref{fig:topos} the market share owned by the top operating system. We observe a behaviour quite similar to the HHI, with a dip around 2011, where the dominant operating system (then Symbian) held roughly 1/3 of the overall market. But both before and after that dip, the market saw instead an operating system getting an absolute majority. Since 2013, the dominant operating system is quite steady around 80\% of the overall market, a situation never attained before, which we can classify as an imbalanced oligopoly (where the imbalance is within the oligopolist circle itself).

\section{Conclusions}
The structure of the mobile operating systems market has been investigated through two concentration indices: the Hirschman-Herfindahl Index and the Four-Firm Concentration Ratio. Their evolution in the 2007-2015 period shows that a tight oligopoly has been a characteristic of this market ever since. A dip in HHI around 2011 marks the passage from the dominance by Symbian to the current dominance by Android. Currently, the four dominant operating systems are installed on over 99\% of new devices. The present oligopoly is however strongly imbalanced, since Android alone holds a share larger than 80\% of the market of new devices.  


\end{document}